\documentclass[manuscript,nonacm]{acmart}

\AtBeginDocument{%
  
}

\setcopyright{none}
\renewcommand\footnotetextcopyrightpermission[1]{}

\begin{document}

\title{His Name, Their Judgment: Expert Authority in Posthumous Persona AI}

\author{Hanjing Shi}
\affiliation{%
  \institution{Lehigh University}
  \city{Bethlehem}
  \state{Pennsylvania}
  \country{USA}
}
\email{hasa23@lehigh.edu}

\author{Dominic DiFranzo}
\affiliation{%
  \institution{Lehigh University}
  \city{Bethlehem}
  \state{Pennsylvania}
  \country{USA}
}
\email{djd219@lehigh.edu}

\begin{abstract}
Persona AI can make deceased experts available for decisions they never encountered. Users may seek these personas precisely because they lack the knowledge needed to judge their advice. We thematically analyze 115 focal Xiaohongshu posts and a nested comment sample concerning unofficial personas of Chinese education advisor Zhang Xuefeng. An installation offer promised expert guidance, while a family's reported use required current records and contextual judgment. Other posts positioned the persona as a questioning aid rather than an expert replacement. Commenters connected continued expertise to updating, permission, and the ability to refuse further work. These findings distinguish access to an expert identity from access to professional help. Keeping the persona current puts living actors in control of what the deceased appears to recommend. Operators therefore need to undertake the support promised by the persona while making their continued use of the person's identity open to challenge.

\end{abstract}



\maketitle

\section{Introduction}

Generative AI is changing what people can ask of the dead. Systems reconstruct loved ones, return performers to public attention, and make deceased people intervene in public disputes \cite{divon2025artificially,meikle2026authority}. Research examines the relationships they sustain and the consent, identity, and commercial interests they unsettle \cite{lei2025afterlife,morris2025ghosts,hollanek2024griefbots}. A further question arises when a reconstruction becomes someone to consult about circumstances that developed after its human source died.

Persona AI presents responses through the recognizable identity, style, or role of a person. Agent skills can make such personas reusable by packaging instructions, reference material, and procedures for compatible systems \cite{agentskills2026}. These packages invite users to apply that person's judgment to new problems. The name can stay the same as developers change the models, tools, and sources producing the answer. An expert's reputation travels with the persona, while their ability to revise or refuse the advice does not.

We call the continuation of a deceased person's social or professional work through new system outputs \emph{functional digital resurrection}. Memory, attachment, and practical assistance can coexist in these encounters. A professional persona also carries expectations tied to the role it claims. Advice about education or employment must address changing opportunities and circumstances specific to the person seeking help. Preserving how an expert once spoke leaves open what would justify speaking for that expert now.

The Zhang Xuefeng case brings this question into consequential family decisions. Zhang helped Chinese high-school students and their families choose universities and majors after the national college entrance examination, and advised on postgraduate study and employment \cite{businessherald2026skill}. These choices shape family expenditure, where students live, and which careers become accessible. After he died on March 24, 2026 \cite{zhong2026obituary}, independent developers drew on his public books, interviews, and speeches to release persona instructions \cite{businessherald2026skill,huashu2026skill}. Posts also described models fine-tuned on broadcast transcripts (post IDs N34 and N58). One title counted ``sixteen days'' since his death (N26). His advisory identity was becoming available for further work while its use was already being contested.

By the 2026 admissions cycle, this possibility had entered one reported family decision. A July Xiaohongshu post described using the coding agent Codex and a Zhang Xuefeng-style skill to prepare university and major preferences for a younger relative (N352). The post author reported purchasing hundreds of pages of admissions records, emphasized checking details, and placed the final decision with the student. Readers asked about extraction errors and offered payment for help, and the post author suggested a way to check against the source PDF. In a separate advertisement, the promise of individual consultation under Zhang's name accompanied a notice limiting the seller's service to installation (N362). An available expert persona could invite expectations of professional help that its operation alone would leave unmet. The promised help would also have to remain useful as admissions requirements changed. A commenter anticipated its expertise becoming obsolete within ``one or two years'' (comment C26.028), while others objected to making the advisor keep working after death. These accounts connected practical help with disputes over who could continue Zhang's advisory work.

We investigate two research questions.

\textbf{RQ1.} How do Xiaohongshu posts and comments frame the help an expert persona can offer and the work needed to make its advice usable?

\textbf{RQ2.} How do these discussions negotiate who may shape, answer for, or end the advisory role performed under a deceased expert's name?

Our analysis identifies a separation between a continuing expert identity and the people doing the work of expertise. Commenters judged the persona against changing knowledge, while the family account located usable advice in records, checking, and contextual judgment. Disputes over consent and posthumous labor questioned who could extend or end the role performed under his name.

We distinguish making an expert identity available from undertaking an advisory service. For someone seeking knowledge they lack, checking an answer may require the professional help the persona promises to provide. Supplying that help involves selecting current information and deciding how the expert's earlier commitments apply to a new situation. When incorporated into the persona, those choices shape what the deceased appears to recommend. Our contribution connects the work of making advice usable with the right to keep advising under a deceased expert's name. Supporting the user and representing the expert are two obligations of the same service.

\section{Related Work}

\subsection{AI resurrection from presence to role performance}

Studies of imagined AI afterlives document expectations of comfort alongside concerns about fidelity, dependence, consent, and control by companies or relatives \cite{lei2025afterlife,morris2025ghosts}. Ethical and legal accounts distinguish access to a person's traces from permission to reconstruct their identity \cite{ohman2018framework,harbinja2023ghostbots,hollanek2024griefbots}. Work in China situates these questions within personality interests, family claims, and commercial grief services \cite{cheng2025law}.

This literature extends beyond private remembrance. Divon and Pentzold develop spectral labor as an account of value produced from the dead \cite{divon2025artificially}. Kidd and Nieto McAvoy examine commercial infrastructures of interactive memory \cite{kidd2025synthetic}. Meikle and Sumiala examine the authority claimed when synthetic videos make deceased people intervene in public disputes \cite{meikle2026authority}.

Our case follows these concerns into repeated requests for expert judgment. Each new question can require an answer beyond the person's recorded words. This makes continuity a question of who can interpret and extend a person's commitments, with current information and permission becoming intertwined in the production of advice.

\subsection{Persona authority and consequential advice}

AI-mediated communication distinguishes the apparent speaker from the system contributing to a message \cite{hancock2020aimc}. Mimetic AI reproduces identifiable people \cite{bukingolts2025mimetic}. A public Steve Jobs skill offers product and strategy advice through rules distilled from biographical material, speeches, and interviews \cite{huashu2026jobs}. Research on self-created text clones shows how people select aspects of themselves and negotiate the clone's limits \cite{lee2026clones}. Posthumous personas leave those choices to others, who decide what will count as the person's continuing judgment.

Lai's study of the \#Keep4o backlash documents practical reliance and emotional attachment developed through interaction with an AI model \cite{lai2026keep4o}. Posthumous personas can also invoke trust associated with a human relationship or public reputation preceding the AI encounter \cite{lei2025afterlife,hollanek2024griefbots}.

Advice research examines perceived expertise, trust, and the sharing of a decision's burden \cite{baines2024advice,gazit2023advisors,aschauer2024advice}. Verification introduces a further difficulty. Virk and Liu found that business professionals struggled to identify flaws in AI-generated analyses despite warnings and incentives \cite{virk2025nonprogrammers}. Three experiments on AI-generated science texts found no consistent main effect of disclaimer type on message credibility \cite{henestrosa2025disclaimers}. Elish's moral crumple zones explain how responsibility can fall on people with limited control over automated systems \cite{elish2019moral}. These studies distinguish willingness to check, capacity to judge, and responsibility for outcomes. Posthumous persona AI adds a question about whose expertise supports reliance when the named advisor can neither oversee the system nor answer for its recommendations.

Digital inequality research distinguishes access and use from the benefits people obtain, linking those benefits to offline resources \cite{vandeursen2015divide}. In a recent preprint, Lee and colleagues estimated greater growth in LLM-assisted admissions-essay writing among applicants receiving fee waivers \cite{lee2026admissionsdivide}. They interpret this pattern as consistent with substitution for scarce writing support.

\subsection{Authority, representation, and avenues for challenge}

Model cards document developers, versions, intended uses, evaluation, and feedback routes \cite{mitchell2019modelcards}, while datasheets address dataset maintenance and contact points \cite{gebru2021datasheets}. C2PA provides verifiable content provenance, distinct from establishing a claim's truth \cite{c2pa2026specification}. Contestable AI connects traceability and explanation with opportunities to challenge decisions over time \cite{alfrink2023contestable}.

For persona AI, these practices intersect with control over representation. Someone can take responsibility for operating a system while their entitlement to speak under its human source's name remains disputed.

China's deep-synthesis and generative-AI rules address labeling, user protection, complaints, and service governance \cite{cac2022deep,cac2023generative}. Labeling measures effective from September 2025 distribute disclosure duties across providers, platforms, and users \cite{cac2025labeling}. Article 994 of the Civil Code provides close relatives a route to protect certain personality interests of deceased people \cite{civilcode2020article994}. After our September 6 post snapshot, Supreme People's Court guidance issued on September 7 addressed AI uses infringing deceased persons' personality interests and providers' duties upon notice of infringement \cite{spc2026ai}. This subsequent guidance informs our discussion of avenues for challenge. Our case asks how permission and responsibility arise within the promise of continued expertise, where a recognizable identity can lend significance to advice that others produce.

\section{Method}

\subsection{Case and platform selection}

Xiaohongshu, also known as RedNote, combines searchable experience-based content with social discussion. Research documents movement between recommendation and search \cite{chen2025qilin}, audience formation through hashtags \cite{wan2025hashtag}, and personal narratives informing responses to fraud \cite{zheng2026scam}. These features support examining how posts present a persona as useful guidance and how readers question that promise.

Weibo leads and web searches supplied contextual news about construction, disclaimers \cite{beike2026skill}, and a company formerly part-owned by Zhang intending to investigate, according to its customer service \cite{guo2026skill}. These attributable sources provide context, separate from the Xiaohongshu corpus.

\subsection{Post discovery and screening}

Using an authenticated Xiaohongshu web session, we collected the post corpus as of September 6, 2026. Six separate queries covered Zhang Xuefeng skills, digital doubles, resurrection, and persona distillation. We recorded the first 100 distinct cards per query in default search order, producing 600 query records and 472 unique posts after deduplication by identifier.

We opened all 472 pages and screened their titles and accessible accompanying text. For image and video posts, analysis used titles and captions. Visible likes guided reading order, rather than eligibility. The screen retained 115 directly related posts about Zhang personas and eight boundary records concerning prediction claims or other memorial representations. We kept 82 contextual records separately, excluded 251 unrelated posts, and set aside 16 with insufficient text. A native Chinese-speaking researcher checked the materials and confirmed relevance decisions.

Appendix~\ref{app:data-dates} details the collection timeline and interpretation of date labels. The supplementary sampling protocol lists all six query strings verbatim with English glosses and provides screening details.

\subsection{Sampling comments within the post corpus}

Post analysis covered all 115 directly related posts and eight boundary records. On September 9, 2026, we visited the comment areas of all 115 directly related posts. Fifty-three contained visible comments, 61 displayed an empty state, and one post was unavailable. A comment thread here means the discussion beneath one post. Top-level comments respond to that post, while replies respond to another comment.

We captured up to 100 top-level comments per post in default display order, retaining initially visible replies linked to their top-level discussion. Two comment areas reached this cap and 51 reached a visible end. The resulting corpus contains 773 top-level records and 159 human replies, with three labeled platform-assistant replies held separately. We preserved displayed date labels and engagement counts alongside observation timestamps.

Fifteen human records contained no captured text. The remaining 917 text-bearing contributions were read in context, with tag-only responses, unrelated text, and promotional asides supplying no thematic evidence. Comment volume varied substantially across posts, so comparisons treated threads as situated discussions rather than weighting themes by comment frequency. Counts describe captured contributions, not independent participants.

\subsection{Analysis and reflexivity}

We used reflexive thematic analysis \cite{braun2019reflexive}. A native Chinese-speaking researcher checked the post screening and interpreted the Chinese post corpus. Comments were read alongside their parent posts and visible replies, with posts as the primary comparison units. Source-linked notes distinguished reported use, demonstrations, advertisements, news, and speculation, and recorded claims about expertise, advisory work, permission, and continued service.

The researcher had a computer science background and had not taken China's national college entrance examination. Relatives sought their technical help to set up a Zhang Xuefeng skill for a younger family member. During use, the researcher found it difficult to assess the soundness of its admissions recommendations. This experience sensitized the reading to the gap between making a persona accessible and judging its advice.

Rereading compared claims within and across posts. Contrasting installation advertising, the family account, and questioning guidance shifted attention from whether a tool was useful to who made it useful. Demands for current knowledge connected technical reconstruction to others' continuing judgment. Accounts of comfort, free help, and useful assistance challenged an initial support-versus-opposition reading. These comparisons developed the four themes below. Themes express interpretive relationships rather than opinion frequencies \cite{mcdonald2019reliability}. Repeated news framings and dense threads remained shared contexts, rather than independent confirmations.

English paraphrases preserve context, while quotation marks identify short translated phrases. N identifies a post's position in the 472-page reading sequence. C identifies a contribution within its parent thread, including replies. Each empirical example is traceable to a saved Chinese source record through a restricted crosswalk.

\subsection{Ethics and data protection}

Under the researcher's institutional policy, this noninteractive analysis of public material fell outside research requiring institutional review. Following social-media research ethics \cite{ford2021ethics}, source text and links remain restricted. Reported examples omit identities and distinctive personal details and use English paraphrase to reduce reverse-search risks. Legal arguments and reported use are interpreted as users' situated accounts.

\section{Findings}

Posts presented a reconstructed Zhang as someone families could consult, while comments questioned what would sustain that role. The first two themes address RQ1 by following the move from reconstructing expertise to making advice usable. The next two address RQ2 through disputes over who may continue that work under Zhang's name, define its tasks, and decide when it should end.

\subsection{What survived of the expert beyond a recognizable voice?}

Reconstruction promised more than familiar speech. N243 presented reusable judgment rules and questions, whereas N58 described a model fine-tuned on transcripts. Instructions condition a host system at use time, while fine-tuning changes model parameters. Both circulated as ``distillation,'' attaching the same expert identity to different ways of producing an answer. N386 also advertised three modes of expression, ranging from blunt advice to gentler comparison and reassurance before recommendations. The post presented this adaptation as a way to address different family circumstances. Its promised persona therefore included choices about how the expert would respond to a student, as well as what information he would provide.

The system behind the name mattered in one reported trial. A user found that their outputs differed from published demonstrations and attributed this partly to the host model (N354). The shared name gave these systems a recognizable advisor, while their answers remained shaped by living people's choices about models, source material, and instructions.

The host-model explanation concerned how answers were generated. Commenters raised a further question about the material available to any reconstruction. What could records of Zhang's past speech preserve of the expertise that produced it? C16.005 distinguished a familiar tone from the advisor's information channels and analytical judgment. A reply immediately added that the information itself changes (C16.006). C26.027 located useful knowledge in institutional contacts and experience that would not necessarily enter a public archive. These comments distinguished preserving an advisor's earlier conclusions from retaining the means by which he reached new ones. One commenter reporting prior use described the system as expressing the user's existing inclination in Zhang's style (C38.046). In that account, sounding like the expert gave the user's own preference the appearance of outside advice.

Preserving earlier judgment also raised the question of how long it would remain useful. Admissions policies and employment prospects could change after the source material was recorded. C26.028 anticipated obsolescence within ``one or two years.'' C26.016 applied the same test to the living advisor, arguing that his public value would have disappeared if he merely repeated last year's advice. C6.026 pointed to changing admissions policies, while C6.041 asked how knowledge would be kept current. These comments treated updating as part of expertise itself. The question was how advice could stay useful once its human source could no longer revise it.

Responses to these expectations also raised questions about how seriously the model should be taken as an advisor. Beneath the fine-tuning demonstration N34, the post author attributed knowledge limits to the base model (C34.105) and proposed retrieval or web search in response to a disputed admission score (C34.107). In a separate exchange responding to hostile criticism, the same author called it ``a toy'' (C34.103). The thread contained both proposed repairs to advisory capability and a description that lowered expectations of it. What users could reasonably expect from the reconstruction remained unsettled.

Other responses identified ways to supplement the archive. C16.089 imagined adding search and agent capabilities, while C26.057 valued a preserved perspective even while expecting limits to its innovation. In the same fine-tuning discussion, a commenter challenged the model's ability to keep up with changing professional prospects (C34.011). The post author proposed collecting other people's experiences and views, converting them into Zhang-style training material, and continuing to fine-tune the model (C34.012). This proposed update would give the persona new material from people other than Zhang. Elsewhere, a commenter listed changing policies, universities, and employment markets and asked who would continually organize this information without payment (C38.075). Keeping the advisor useful thus raised two connected questions about who would supply current knowledge and whose judgment it would become under his name. The family account below shows how additional work entered an admissions decision.

\subsection{From installing a persona to obtaining advice}

Posts offered access to Zhang's advice through a free skill (N250) and a website usable on phones or computers (N321). A free persona could still prompt further spending or require help to use. The user who compared host models described considering a second model subscription after finding the results unlike the demonstration (N354). In one news account, a journalist relied on professional assistance to install the software (N74). An advertisement turned this technical step into an offer of expert access (N362). It appealed to parents' lack of admissions knowledge and dissatisfaction with expensive agencies, promising individual consultation with Zhang at home. The advertised service included remote installation, setup, and instruction, but ended, ``I only provide installation services,'' leaving decisions with students and parents. The advertisement used parents' need for guidance to promote a service whose stated responsibility ended with technical setup. Obtaining the tool and obtaining help judging its advice were different services.

Readers also sought help from people offering access. Beneath a web-based persona, a reader reported problems and the post author invited them to send details (C291.001--C291.002). Under the free admissions assistant, a question about reliability received reassurance and an invitation to try it (C321.001--C321.002). These responses offered contact or encouragement, while leaving the assessment of particular recommendations open. What further help would a family need once the tool was working?

A separate July 2026 account described what one family did after gaining access (N352). The post was written by the student's older sibling, who described a household unfamiliar with admissions procedures and with limited connections to relevant industries. The sibling purchased the current year's admissions plans and three years of historical records, then combined Codex with admissions and Zhang Xuefeng-style skills. The two documents contained nearly 600 scanned pages. According to the post, Codex extracted fields such as program codes, tuition, and prior admission scores and organized candidate lists around different priorities. One prioritized employment and value, while another incorporated the student's interests and possible further study.

For the sibling, the benefit was discovering options the family could discuss. Programs in areas such as supply-chain and media operations broadened a search previously centered on a few familiar choices. The post attributed this benefit to the combined workflow, with the persona contributing a way to approach the decision. Its useful heuristics included weighing tuition and the cost of a wrong turn, considering ordinary graduates' prospects, and working backward from possible employment. The family then tested these priorities against the student's interests, subject strengths, and the physical demands of possible occupations. Strong employment prospects alone were insufficient grounds for recommending a program. The account placed the final choice with the student, describing how the family developed options rather than reporting an admissions outcome.

This useful expansion of options coexisted with errors in the material used to compare them. The sibling reported data discrepancies and stressed checking details. One reader who had also tried extracting admissions tables asked, ``How can this be guaranteed?'' (C352.001). In reply, the sibling suggested converting the tables into readable form and checking them against the source PDF (C352.002). That response offered a checking procedure for the reader to carry out. In a separate exchange, another reader asked, ``Could you help me with this for a small payment?'' (C352.003). The sibling invited private contact, although any subsequent transaction remains unknown (C352.004). The exchanges show assistance continuing around the tool, through requests to the person who had demonstrated its use.

The family account also clarifies the distinction between listing programs and deciding what makes a program suitable. Comments in another discussion raised that distinction explicitly. C26.031 declined to trust the system with program selection. C26.061 contrasted answering how to learn programming with first asking whether someone should pursue it. A fluent answer could address the stated question while leaving the decision behind it unexplored.

A separate guide made exploring that decision the persona's purpose (N349). It recommended using the skill as a questioning aid before asking it to choose programs. The suggested sequence first elicited budget, geography, unacceptable risks, and future study plans, then identified candidate options, missing information, and matters requiring verification. Official admissions sources supplied the next checks. The guide's author also offered to help someone try it, prompting a reader to ask whether that help remained available (C349.001--C349.002). This was an offer of assistance, distinct from the family's reported experience. Where the advertisement promised access to an advisor, the guide specified help in working out what to ask and check. In the family account, such work was supplied through purchased records, the sibling's preparation, and discussion of the student's circumstances. Access to generated advice opened the decision process, while these additional contributions made the options usable.

\subsection{Publicly available words became a dispute over future speech}

The preceding accounts located usable help in work supplied by people around the persona. When new information and judgments enter the persona's responses, they also change what Zhang appears to advise. Readers disputed whether helping present-day users entitled others to make those choices under his name. N165 asked whether the family or company had approved third-party reconstruction and challenged the idea that free distribution meant the creators had nothing to gain. It pointed to attention and audience growth as possible benefits. N174 proposed routes for relatives and the company to challenge reconstruction, while N308 disputed how legal rights were being described. The speed of reconstruction made the question of permission more immediate in these accounts. N26 counted sixteen days since death, and N130 contrasted his recent departure with an AI version becoming available. The ability to release a persona quickly had become part of the dispute over whether others should do so.

Defenses of continued use appealed to what it could offer the living. C4.001 reasoned that a free tool helping more people could accord with Zhang's commitments. Beneath a post opposing his reconstruction, C133.001 emphasized help for ordinary people. C133.005 went further, framing opposition to resurrection as opposition to families with fewer resources. In that argument, restricting reuse meant withholding needed help, making access a reason to defend continuation itself. C21.085 defended open-source distribution as noncommercial. C26.079 offered an analogy to ordinary learning, asking how studying his books and then advising others would differ. In this account, a persona extended the circulation of knowledge already shared in public. C16.028 imagined a different benefit, with an authorized reconstruction offering comfort to relatives. These defenses drew on access, learning, and remembered relationships to explain why continuation could be valued.

The difficulty was deciding who could turn those possible benefits into permission. C4.011 accepted continuation if it fulfilled Zhang's wishes or had family support, while C4.071 challenged commenters who assumed his consent. C26.024 asked whether income from distilled ability should benefit descendants. The issue also extended beyond reuse of existing words. C21.027 worried that listeners might accept statements Zhang would never have made. Read against the learning analogy, this concern distinguished drawing on an expert's public work from issuing new answers under his identity. The dispute concerned who could invoke Zhang's remembered wishes to authorize that future speech, and whose interests it would serve.

\subsection{From remembrance to work without an exit}

The permission dispute extended from what the persona could say to how long it could be made to serve. N369 contrasted Zhang's career helping others make choices with his inability to choose whether to become an AI. This reversal placed the represented person's agency at the center of the issue. N293 carried the tension into an imagined workplace. A colleague who had left the job could be reconstructed from messages and documents to keep writing code and replying to others. The post asked who would be responsible for this digital employee's mistakes. The scenario concerned a former employee, extending the question beyond death to whether leaving a workplace could end work performed under one's name.

C16.003 condemned profiting from another's suffering through the Chinese metaphor of a ``human-blood bun.'' The commenter contrasted burial with being kept at work for the living, distinguishing knowledge that outlives someone from continued demands on that person's identity. C16.039 described people becoming production material, while C26.026 objected to preserving only someone's working identity. What these comments resisted was the reduction of a person to useful outputs that others could keep requesting. C16.064 offered the joke of disconnecting the power. Switching off one system provided an imagined endpoint, while the dispute concerned who could end further uses of a reusable persona.

The ability to keep a persona working also lets others decide what work it should do. A reusable persona could be assigned tasks beyond the advisory role it recalled. Four boundary captions circulated warnings about videos using an AI Zhang Xuefeng to predict examination essay topics (N23, N59, N159, N196). These were critical reports of prediction claims, rather than observed use of the sampled skills. Predicting an unpublished examination question called for a different basis than comparing programs using admissions records. The expert's recognizable identity could travel between these tasks even when their evidentiary demands differed. The captions challenged that extension, while the imagined employee raised the parallel question of who could assign further work after departure. In both, keeping a persona available also gave living people opportunities to redefine its job. The disputes therefore concerned what the persona would know, whom it would serve, and who could decide when that service should end.

\section{Discussion}

The case reveals a split between the person who makes advice worth seeking and the people who make it usable. Installation, source checking, and judgment about personal fit are different undertakings, even when one expert name connects them. This distinction raises linked questions about whose judgment speaks through the expert, what support users can obtain, and who can limit continued service.

\subsection{A person can appear to continue while others take over the judgment}

A professional reputation comes from judgments made over a changing career. An expert persona uses that reputation to offer guidance, as the installation advertisement did with Zhang's name. The name identifies whom users are invited to consult, while others choose the models and sources shaping the advice. Keeping the archive unchanged risks outdated advice. Updating it introduces choices about which new information matters and how earlier commitments apply. When others' experiences are rewritten as the expert's training material, maintaining usefulness also becomes an act of representing that person. Those choices shape what the deceased now appears to recommend.

The report of preferences restated in Zhang's style suggests that a persona may also give a user's own inclination the appearance of independent advice. For someone seeking a perspective beyond their own, factual accuracy alone would leave open whether the persona offers grounds to reconsider a choice or simply lends an expert identity to it.

Digital-afterlife research already examines synthetic public authority \cite{meikle2026authority} and permission to update evolving agents \cite{lei2025afterlife}. Our case places these issues inside an advisory relationship with strangers. Its distinctive problem is that making the service useful requires extending the deceased's judgment, even while the right to make that extension remains disputed. A living advisor could explain why their recommendation changed. With a posthumous persona, that explanation must come from someone else, whose authority to speak for the advisor may itself be contested. A correct answer and an authorized representation become separate obligations.

This separation changes how continuity should be understood (Figure~\ref{fig:continuation}). Self-created clones leave editorial choices partly with the represented person \cite{lee2026clones}, whereas posthumous expert personas put those choices elsewhere. In Lai's \#Keep4o study, a model change disrupted relationships formed through interaction \cite{lai2026keep4o}. Here the expert's name identifies a familiar advisor while leaving unclear whose judgments now shape his recommendations. The person appears to continue precisely when others have taken over the decisions about what he would say.

\begin{figure}[tb]
  \centering
  \begin{tabular}{p{0.43\linewidth}p{0.49\linewidth}}
    \toprule
    \textbf{What the persona carries forward} & \textbf{What remains contested} \\
    \midrule
    \textbf{His name} & \textbf{Their judgment} \\
    Reputation and remembered commitments & Others select models, sources, and new answers \\[0.7em]
    \textbf{Expertise on demand} & \textbf{Work needed to make advice usable} \\
    A reusable advisory persona & Families prepare records and check suitability \\[0.7em]
    \textbf{Continued availability} & \textbf{Permission to continue or stop} \\
    New tasks performed under the same identity & The represented person can no longer refuse \\
    \bottomrule
  \end{tabular}
  \caption{Three tensions in continued expertise. Identity can persist across changes in judgment, an available persona can require further advisory work, and continued service can exceed the person's ability to refuse. These are interpretive relationships drawn from the case.}
  \Description{A two-column comparison connects what a persona carries forward with what remains contested. A retained name contrasts with others selecting models and sources. Expertise on demand contrasts with families preparing and checking records. Continued availability contrasts with contested permission when the represented person can no longer refuse.}
  \label{fig:continuation}
\end{figure}

\subsection{The promise of accessible expertise can relocate its burdens}

The promise of a freely reusable expert enters a market where professional help has a price. A 2024 report listed admissions packages associated with Zhang's service at 11,999 and 17,999 yuan, with limited places remaining before the examination \cite{lanjing2024admissions}. The contrast between an installation offer, a family account, and questioning guidance locates a further threshold after access. Preparing evidence and judging suitability still require someone to undertake the work that makes advice useful.

Advice can help people identify options and decide how to evaluate them \cite{baines2024advice}. A student unfamiliar with a field may need help recognizing which differences between programs matter. Checking an admission score leaves open whether suitable programs were omitted or employment prospects were given too much weight. Virk and Liu found that even motivated business users struggled to assess AI-generated analyses \cite{virk2025nonprogrammers}. Their study distinguishes encouraging scrutiny from enabling it. For expert personas, the corresponding question is who helps users judge the options they needed assistance to identify.

Digital inequality research distinguishes obtaining technology from benefiting from it \cite{vandeursen2015divide}. In the family account, AI brought unfamiliar programs into view, while purchased records and relatives' judgment helped assess them. Discovering more options and needing help to judge them can coexist. Requests for paid help and interest in additional model subscriptions suggest that obtaining a persona can leave further resource needs unresolved. A persona might supplement professional advice for one household and substitute for unavailable advice for another. Asking the latter to consult an expert could return them to the resource constraint that brought them to the persona.

Elish's moral crumple zones explain responsibility falling on people with limited control \cite{elish2019moral}. In the installation advertisement, Zhang's name invited reliance on professional guidance, while the seller committed only to installation. The disclaimer records a service boundary, rather than establishing legal immunity. The fine-tuning discussion adds uncertainty about the service's scope, as its creator proposes improvements in some exchanges and calls the model a toy in another. A contemporary interview similarly assigned designers responsibility for gaps between persona marketing and capability \cite{luo2026digitaldouble}. The mismatch arises before any demonstrated harm. If checking the recommendation requires the expertise sought, leaving the decision to the user also leaves unresolved who will help them assess it.

\subsection{A reusable identity makes the boundaries of work contestable}

Support for users leaves another obligation unresolved, toward the person whose name makes the service recognizable. Spectral labor explains how synthetic afterlives produce value from the dead \cite{divon2025artificially}. The workplace analogy in our corpus adds a question about withholding future work. Once a professional identity becomes reusable instructions, death or departure can cease to be a practical stopping point for outputs issued under that identity. Continued technical availability then exceeds the person's opportunity to accept another assignment.

Task boundaries matter as much as duration. The examination-prediction warnings challenged claims made under Zhang's name about future essay topics. The disputed extension was from advising on admissions to predicting unpublished questions under the same expert identity. The question is who is entitled to assign such tasks to a reconstructed person.

Permission therefore needs a scope and a duration. Approval of reconstruction, a particular task, and indefinite future service are different commitments. The access-based defenses make this distinction difficult in practice, because restricting representation can be framed as denying help to people who need it. That argument links the interests of prospective users to decisions about another person's identity, without resolving who may make those decisions. Design must accommodate requests for assistance and challenges to the continued use of the identity that makes such assistance attractive.

\subsection{Making operators responsible for the work behind the name}

An expert-persona service needs to help users assess advice and make its use of another person's identity open to challenge. We use \emph{role provenance} to connect these obligations to identifiable operators, extending documentation and contestability practices \cite{mitchell2019modelcards,gebru2021datasheets,alfrink2023contestable}.

The Supreme People's Court guidance described above illustrates a route for challenging representation through claims about a deceased person's name, likeness, or reputation \cite{spc2026ai}. Such claims address injury to the represented person. Our findings identify a corresponding need on the user's side, for help recognizing unsuitable recommendations when doing so requires the expertise sought. Role provenance places these two forms of support within the operation of the service.

Responsibilities differ with control. Skill authors shape instructions, model providers shape underlying capabilities, and operators decide how the combination reaches users. An admissions interface could distinguish installation, question preparation, and qualified advice review before a family begins. For reviewed advice, it could show the admissions cycle, the reviewer responsible, and unresolved constraints beside each recommendation. A request for help would route those constraints and source records to that reviewer, who could return a correction or explain what further information was needed. Changes to the model or source records would trigger renewed review. A dated review commitment assigns work, rather than certifying advice until an expiry date.

Such support costs resources, and free skills may lack a willing reviewer. Question preparation offers a bounded alternative in that situation. A separate route for challenging representation should let legitimate representatives request restrictions on the persona's tasks or continued availability. These proposed interactions require evaluation of whether help reaches users and restrictions take effect, as well as whether the arrangements are affordable.

\section{Limitations}

This single-platform case captures public accounts of a newly circulating persona. Personalized search and fixed discovery depth shape visibility, while title-and-caption analysis leaves audiovisual meanings outside its scope. Although every retained focal comment area was visited, unequal comment volume and default display order make some discussions more visible than others. One post was unavailable and two reached the capture cap. The detailed family account supports a situated comparison with service promises and guidance, rather than a general account of household practice.

Public accounts provide limited access to how advice was used outside the platform. The study does not establish adoption rates, household resources, decision outcomes, or differences in checking capacity between social groups. Interviews and observation could follow how families identify alternatives, obtain assistance, and decide when to question a recommendation.

Comparative studies could examine other professional and intimate personas, including how different models change advice issued under the same name. The reported experience of having one's preferences restated in an expert's style also calls for testing when persona advice broadens a user's perspective and when it reinforces existing inclinations. Future work should test whether operator commitments reduce users' checking burden while enabling meaningful limits on continued representation.

\section{Conclusion}

An expert's name can survive while the work of judgment passes to others. The Zhang Xuefeng discussions trace this separation through demands for current knowledge, family checking, and objections to service without refusal. Preserving a recognizable advisor and providing professional help are different achievements. Making advice useful requires people to interpret records and personal circumstances. What they incorporate into the persona also shapes what the deceased appears to recommend.

Evaluation must therefore ask whether users can obtain help judging advice, alongside who may continue speaking under the deceased's name. This single-platform study motivates observation of advisory use and evaluation of provider commitments that support decisions while allowing representation to be challenged.

\bibliographystyle{ACM-Reference-Format}
\bibliography{refs}

\appendix
\section{Data Collection Dates}
\label{app:data-dates}

Post discovery and screening used a September 6, 2026 snapshot. The comment areas of the retained focal posts were visited on September 9. Both dates use the researcher's local time (UTC$-$07:00), corresponding to some September 7 and September 10 observation timestamps in UTC, respectively. Observation timestamps record when material was accessed, separately from the publication or edit labels displayed by the platform.

Among the 115 directly related posts and eight boundary records, 120 displayed month-day publication or edit labels spanning March 24 through August 17. One displayed the relative label ``two days ago,'' and two lacked a captured date label. Where the interface omitted the year, 2026 was inferred from its current-year convention. Relative labels were retained as displayed rather than converted into precise dates. Labels identifying an edit were kept distinct from original publication dates.

Among captured human comments and replies, month-day labels ranged from March 24 through August 12, with the year interpreted using the same convention. Two contributions displayed the relative label ``four days ago.'' The comment observations describe what was accessible on September 9, rather than establishing that every contribution was visible at the earlier post-collection date. Seven threads contributed 591 of the 773 top-level records. The remaining 46 nonempty threads supplied 182, making the uneven distribution visible alongside coverage of the full focal post set.

\end{document}